\documentclass[pra,amsmath,twocolumn,showpacs]{revtex4}
\usepackage{graphics}

\begin{document}
\title {Slow Light Propagation in a Thin Optical Fiber via Electromagnetically Induced Transparency}
\author {Anil K. Patnaik}
\affiliation{Department of Applied Physics and Chemistry, University of Electro-Communications, Chofu, Tokyo 182-8585 Japan\\
CREST, Japan Science and Technology Corporation, Chofu, Tokyo 182-8585 Japan}
\author {J. Q. Liang}
\affiliation{Department of Applied Physics and Chemistry, University of Electro-Communications, Chofu, Tokyo 182-8585 Japan\\ 
CREST, Japan Science and Technology Corporation, Chofu, Tokyo 182-8585 Japan}
\author{K. Hakuta}
\affiliation{Department of Applied Physics and Chemistry, University of Electro-Communications, Chofu, Tokyo 182-8585 Japan\\
CREST, Japan Science and Technology Corporation, Chofu, Tokyo 182-8585 Japan}
\date{June 17, 2002}
%\date{\today}
 
\begin{abstract}
We propose a novel configuration that utilizes electromagnetically induced transparency (EIT) to {\em tailor} a fiber mode propagating inside a thin optical fiber and coherently control its dispersion properties to drastically reduce the group velocity of the fiber mode. The key to this proposal is: the evanescent-like field of the thin fiber strongly couples with the surrounding active medium, so that the EIT condition is met by the medium. We show how the properties of the fiber mode are modified due to the EIT medium, both numerically and analytically. We demonstrate that the group velocity of the new modified fiber mode can be drastically reduced  ($\approx$44 m/sec) using the coherently prepared orthohydrogen doped in a matrix of parahydrogen crystal as the EIT medium.
\end{abstract}

\maketitle

\section{Introduction}

There has been many excitements on the group velocity management via electromagnetically induced transparency to drastically reduce the group velocity of light \cite{slow_review,slowGV,slow_hot,slow_solid}, to freeze \cite{freezeL_exp}, to
store the quantum information of light \cite{store_QL}, to achieve superluminal group velocity propagation \cite{super_L}, and to control the group velocity between subluminal to superluminal propagation \cite{sub_super_L} in varieties of resonant mediums. Many experiments have been reported on reduction of group velocity of light (slow light) using ultra-cold atoms \cite{slowGV}, in hot atoms \cite{slow_hot} and also recently in a rare-earth ion doped crystal \cite{slow_solid}. Slow light has also been observed in far-off-resonance Raman systems \cite{slow_raman}. Extensive study of the slow light can be found in many recent literatures \cite{slow_review} including many interesting applications \cite{slow_NLO,slow_appl_AO,slow_appl1,slow_appl2}. Harris and coworker have demonstrated possibility of observing non-linear processes at low light level using the slow light \cite{slow_NLO}. Matsko {\em et al.} have demonstrated enhancement of acousto-optical effects in a dielectric fiber doped with resonant three-level system when velocity of light becomes comparable to velocity of sound \cite{slow_appl_AO}. Possible applications of slow light for efficient four-wave mixing \cite{slow_appl1} and polarization splitting \cite{slow_appl2} are also reported. It has also been observed that dynamically turning ``off" and then turning ``on" the strong control field, a signal pulse can also be freezed and retrieved \cite{freezeL_exp, store_QL}, leading to possible realization of a quantum storage device. The key to all the above phenomena is existence of large dispersion behavior combined with the very low absorption due to EIT \cite{EIT}. The most attractive feature of the EIT mechanism has been the dynamical control of the dispersive property of the medium using the control laser parameters.

	In recent years rapid progress has also been made in slow light propagation in photonic heterostructures. It is reported that extremely slow light can be observed in photonic band gap materials \cite{slow_PBG}, in Moir\`{e} fiber Bragg gratings \cite{slow_MFG}, and in optical resonator-array waveguides \cite{slow_resonator}. In these kind of studies the material dispersion, which is determined by the engineered structure, is used to achieve the slow light. These materials have significant importance due to their applications to scalable photonic devices. However the structure dependent dispersion-management lack in the dynamical control compared to the EIT assisted control of dispersion.

	In a conventional optical fiber, a propagating single-mode field experiences a dispersion due to the dispersive material contents of the core and clad. The group velocity of the envelope of such a propagating fiber mode is given by
\begin{equation}
v_g = \left[ \frac{d\beta}{d\omega} \right]^{-1},
\label{vg_def}
\end{equation}
where $\beta$ is the propagation constant of the fiber mode and $\omega$ is the angular frequency. In conventional optical fibers such dispersion is very small compared to the dispersions in fibers using {\em tailored} materials such as photonic heterostructures \cite{slow_PBG,slow_MFG,slow_resonator}. 

\begin{figure}
\includegraphics{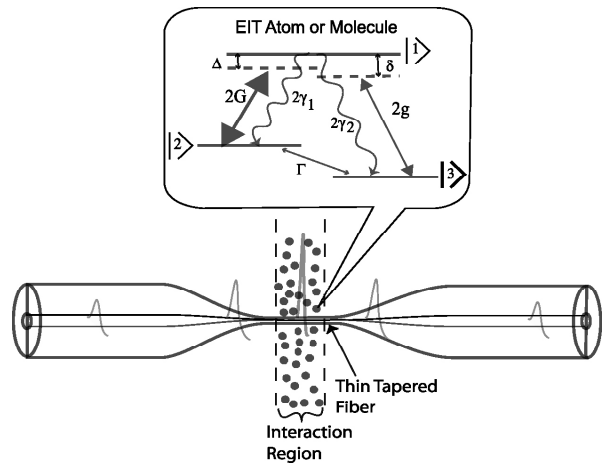}
\caption{
\label{Fig1}
The configuration under consideration: a thin tapered optical fiber is surrounded by the EIT atoms or molecules. Two fields, a strong and a weak field, propagate through the single-mode thin optical fiber. The evanescent-like part of the fiber modes interact with the EIT atoms/molecules within the interaction region as shown above. The energy-level scheme for the EIT atoms/molecules is shown in the inset. The spontaneous decay coefficients are denoted as $2\gamma_i$, the strong control field and weak probe field Rabi frequencies are given by $2G$ and $2g$ respectively, their corresponding frequency detunings, from the transitions they couple, are $\Delta$ and $\delta$. The ground state dephasing is denoted as $\Gamma$. More details of the EIT scheme is described in Sec.\ III.}
\end{figure}

	In this paper we propose a configuration to {\em tailor} the dispersion properties of a fiber mode propagating inside a thin (tapered) optical fiber surrounded by an EIT medium [see Fig.\ \ref{Fig1}], and as a result we show that the group velocity of the fiber mode can be reduced significantly. Our method utilizes EIT and hence has the advantage of large control over the dispersion property of the fiber mode. It should be noted that our usage of ``thin fiber" should be understood as the case when the diameter of the fiber $2a \stackrel{<}{\scriptstyle\sim} \lambda$, $\lambda$ is the central wavelength of the field that propagates inside the fiber. Under such a condition large part of the energy (evanescent-like) of the fiber mode lies outside the thin fiber.  Thin tapered fiber has been a standard method in the dielectric sphere experiments to get large coupling of the output field with the whispering-gallery modes of the sphere \cite{sphere}.

	Further we have introduced ortho-H$_2$ doped in a matrix of para-H$_2$ as an EIT medium surrounding the fiber to demonstrate the above mentioned reduction in $v_g$. This medium has the following advantages over other EIT mediums: (a) It has very high density and short interaction length (few tens of $\mu$m) compared to its gas phase. (b) The vibrational states considered for the EIT transitions have very small dephasing ($\sim$53 Hz at an operating temperature of 4K \cite{Gamma_dph}). To our knowledge, the only other solid in which EIT has been reported is Pr:YSO crystal \cite{slow_solid,EIT_solid}. Using our scheme we achieve a large dispersion of $\beta$ and hence a huge reduction in the group velocity ($\sim$44 m/sec for a chosen set of field parameters). We also predict the possibility of stopping the fiber mode. With the rich technologies available for the optical fibers, we hope, slow light in optical fiber could open up new possibilities in non-linear optics studies and for integrated photonic device applications.

	The organization of the paper is the following: In Sec.\ II, we present the propagation of a fiber mode inside a thin single-mode optical fiber, particularly when the medium surrounding the thin fiber is an active medium. In Sec.\ III, we describe how the behavior of the fiber mode can be modified when the surrounding medium is an EIT medium and introduce a dressed fiber mode. In Sec.\ IV, we obtain an approximate analytical expression for the group velocity of the new modified fiber mode and show how the group velocity can be substantially reduced by choosing the appropriate parameters. In Sec.\ V, we give a detailed account of realization of slow group velocity of the fiber mode using ortho-H$_2$ doped parahydrogen crystal as an EIT medium surrounding the fiber. We present more detailed aspects of the slow light propagation of the dressed fiber mode with numerical results in Sec.\ VI. We summarize our results in Sec.\ VII.

\section{Propagation Dynamics inside a thin fiber surrounded by an active medium}

	A thin fiber is obtained by heating a conventional communication fiber and pulled to obtain a narrow waist \cite{tap_F}. The waist of thin fiber is $2a \stackrel{<}{\scriptstyle\sim} \lambda$. Thus it has almost a vanishing core and hence the refractive indices that determine the fiber modes are the cladding refractive index which we denote as $n_f$ (and will be referred as fiber refractive index henceforth) and the medium refractive index $n_m$ surrounding the thin fiber. 

	The general equation for propagation of an electric field $\vec{E}_\alpha$ through a medium with a refractive index $n$ is governed by
\begin{equation}
\nabla \times (\nabla \times \vec{E}_\alpha) 
+\frac{1}{c^2}\frac{\partial^2(n^2\vec{E}_\alpha )}{\partial t^2}
= 0,
\label{gen_eq}
\end{equation}
where, $c$ is the velocity of light in free space, and  $\alpha$ is a field index that distinguishes the fields, when more than one laser field propagate through the fiber. Let us consider propagation of a pulse inside the thin single-mode fiber. A pulse corresponding to the lowest order characteristic mode of the fiber with central frequency $\omega_\alpha$ may be given by
\begin{equation}
\vec{E}_\alpha (r,z;t) = \hat{{\cal E}}_\alpha 
\int_{-\infty}^{\infty} {\cal E}_\alpha (r;\omega) e^{i\beta_\alpha z-i\omega t} d\omega + {\rm c.c.}
\label{pulse}
\end{equation}
Here ${\cal E}_\alpha (r;\omega)$ is the distribution of the field amplitude at frequency $\omega$ in the transverse direction and $\beta_\alpha$ corresponds to the propagation constant of the propagating mode with $\omega_\alpha$ being the central frequency. On substituting (\ref{pulse}) into (\ref{gen_eq}) the equation for the field amplitude is obtained as
\begin{equation}
\frac{\partial^2{{\cal E}}_\alpha}{\partial x^2}
+ \frac{\partial^2{{\cal E}}_\alpha}{\partial y^2} 
+ (n^2 k_\alpha^2 - \beta_\alpha^2) {{\cal E}}_\alpha
= 0,
\label{intem_1}
\end{equation}
where $k_\alpha = \omega_\alpha /c$ is the free space propagation constant corresponding to $\omega_\alpha$. The refractive index $n = n_f~(n_m)$ inside (outside) the fiber. In deriving Eq. (\ref{intem_1}), we have made the following approximations: (a) The spatial variation of the refractive index along the transverse direction is assumed to be zero inside the fiber and is negligible outside the fiber (i.e. $\nabla n \simeq 0$).
(b) The time-duration of the pulse is assumed to be long compared to the response of the medium, so that the medium can be considered to be in steady state ($\partial n_m/\partial t = 0$). (c) Further we work in the single (lowest order) mode of the fiber - i.e. under the condition \cite{marcuse}
\begin{equation}
\lambda_c = \frac{2\pi a}{\zeta_c} \sqrt{n_f^2 - n_m^2} < \lambda_\alpha.
\label{single_mode}
\end{equation}
Here $\lambda_\alpha = 2\pi /k_\alpha$, $a$ is radius of the thin fiber, $\zeta_c = 1.405$ corresponds to the first zero of the Bessel function $J_0(\zeta_c) = 0$. (d) We also use the weakly guiding approximation ($(n_f - n_m)/n_m \ll 1$) for which the approximate solution of Eq.\ (\ref{intem_1}) corresponding to the lowest-order mode $HE_{11}$ is a linearly polarized LP$_{01}$ mode \cite{WGF}.

	The propagation equation (\ref{intem_1}) for fiber has well-known analytical solution for the fundamental mode \cite{sphere,marcuse}
\begin{subequations}
\label{fiber_mode}
\begin{eqnarray}
{\cal E}_\alpha (r) &=& A_\alpha\frac{\kappa_{m_\alpha} J_0 (\kappa_{f_\alpha} r)}{\varphi_\alpha {\sqrt \pi}J_1(\kappa_{f_\alpha}a)} ~ {\rm for}~r\le a
\label{fib_modeA} \\
&=& A_\alpha\frac{\kappa_{m_\alpha} J_0 (\kappa_{f_\alpha} a)}{\varphi_\alpha {\sqrt \pi} J_1(\kappa_{f_\alpha} a)} e^{-\phi_\alpha (r-a)}~{\rm for} ~r>a;
\nonumber \\
\label{fib_modeB}
\end{eqnarray}
\end{subequations}
which is obtained by matching the boundary conditions for normal and tangential components of the field separately and is normalized. Here $r = \sqrt{x^2 + y^2}$ is the radial distance from the axis of the fiber, $a$ is the radius of the thin fiber and $A_\alpha$ is a constant that determines the amplitude of the field, and
\begin{eqnarray}
\kappa_{f_\alpha} = \sqrt {k_\alpha^2n_f^2 -\beta_\alpha^2}&,&~\kappa_{m_\alpha} = \sqrt {\beta_\alpha^2 - k_\alpha^2n_m^2}
\nonumber \\
\varphi_\alpha = k_\alpha a\sqrt {n_f^2 - n_m^2}&,&~ \phi_\alpha = \kappa_{m_\alpha} \frac{K_1(\kappa_{m_\alpha}a)}{K_0(\kappa_{m_\alpha}a)}.
\label{fib_coeff}
\end{eqnarray}
The symbols $J_l$ and $K_l$ represent Bessel functions of first kind and modified Bessel functions respectively of order $l$. In writing the above solution the field outside the fiber is approximated to an exponential \cite{sphere} for ease of further evaluation of the integrals. The characteristic equation that determines the propagation constant $\beta_\alpha$ for the fundamental mode is
\begin{equation}
\kappa_{f_\alpha} \frac{J_1(\kappa_{f_\alpha} a)}{J_0 (\kappa_{f_\alpha} a)} = \kappa_{m_\alpha} \frac{K_1 (\kappa_{m_\alpha} a)}{K_0 (\kappa_{m_\alpha} a)}.
\label{beta_eq}
\end{equation}

	The radial distribution of the energy of a fiber mode propagating inside the thin fiber, given by Eq. (\ref{fiber_mode}), clearly depends on the frequency of the propagating field, fiber parameters such as diameter of the fiber ($2a$), refractive index $n_f$, and refractive index of the medium $n_m$. For example, for $k_\alpha a = 1.31$ and $n_f = 1.43$, about $49\%$ of the total energy of the fiber mode lies out side the thin fiber when refractive index of the medium is $n_m = 1$.

	Generally the medium outside the fiber is air or a passive medium having  constant refractive index $n_m$, for which the above mode calculation is valid. But in the present case under consideration, we have an active medium that exists surrounding the fiber, where the refractive index of the medium $n_m \equiv n_m (r)$ is no more a constant,  moreover it could be a function of parameters of the propagating field. In such a case the above description of mode solution fails. However we use an indirect and approximate approach to determine $\beta_\alpha$ using an effective refractive index for the medium to replace $n_m$, which is described as follows:

	Let us introduce an effective refractive index $\bar{n}_m$ for the active medium outside the thin fiber, experienced by the field $\vec{{\cal E}}_p$. The corresponding polarization of the medium at a radial distance $r$ is given by
\begin{equation}
\vec{P}_{\alpha_{av}} (r;\omega_\alpha) = \epsilon_0 (\bar{n}_m^2 -1) \vec{{\cal E}}_\alpha (r;\omega_\alpha).
\label{av_pol}
\end{equation}
Here $\epsilon_0$ is the electrical permittivity of the vacuum. However the actual polarization of the medium has a radial distribution, given by
\begin{equation}
\vec{P}_{\alpha_m} (r;\omega_\alpha) = \epsilon_0 (n_m^2(r) -1) \vec{{\cal E}}_\alpha (r;\omega_\alpha).
\label{pol}
\end{equation}
For $\bar{n}_m$ and hence $\vec{P}_{\alpha_{av}}$ to represent the actual physical system described by Eq.\ (\ref{pol}), we demand that the total interaction energy densities $\vec{{\cal E}}_\alpha^*\cdot\vec{P}_{\alpha_{av}}$ and $\vec{{\cal E}}_\alpha^*\cdot\vec{P}_{\alpha_m}$ over the whole transverse plane due to $\vec{P}_{\alpha_{av}}$ and $\vec{P}_{\alpha_m}$, respectively, should be the same. Thus multiplying $\vec{{\cal E}}_\alpha^* (r;\omega_\alpha)$ to both Eqs.\ (\ref{av_pol}) and (\ref{pol}), integrating over the transverse direction, and equating their right hand sides we get 
\begin{equation}
\bar{n}_m^2 = \frac{\int_a^R {\cal E}_\alpha^*(r) n_m^2 (r) {\cal E}_\alpha(r) r dr}{\int_a^R |{\cal E}_\alpha (r)|^2 r dr}
\label{n_av}
\end{equation}
in cylindrically symmetric coordinates. Here $R$ is the radial extent of the medium from the axis of the fiber. Different schemes for averaging refractive index are used to determine the field propagation in heterogeneous structures (e.g. see in \cite{BPM_num}). 

	Therefore $n_m$ in Eqs.\ (\ref{fiber_mode}-\ref{beta_eq}) can be replaced by the averaged refractive index of the medium $\bar{n}_m$ that is constant along the radial  direction. Thus parameters in Eq.\ (\ref{fib_coeff}) are modified to 
\begin{eqnarray}
\kappa_{f_\alpha} = \sqrt {k_\alpha^2n_f^2 -\beta_\alpha^2}&,&~\kappa_{m_\alpha} = \sqrt {\beta_\alpha^2 - k_\alpha^2\bar{n}_m^2},
\nonumber \\
\varphi_\alpha = k_\alpha a\sqrt {n_f^2 - \bar{n}_m^2}&,&~\phi_\alpha = \kappa_{m_\alpha}\frac{K_1(\kappa_{m_\alpha}a)}{K_0(\kappa_{m_\alpha}a)},
\label{coeff_av}
\end{eqnarray}
that determine the field distribution ${\cal E}_\alpha (r)$ and the propagation constant $\beta_\alpha$ of the fundamental mode. However it should be noted that ${\bar n}_m(r)$ in Eq.\ (\ref{n_av}) itself is a function of ${\cal E}_\alpha (r)$. Therefore, the propagation is initiated, for example, with a Gaussian field at the input, and soon that evolves to the fundamental mode of the fiber ${\cal E}_\alpha (r)$ given in Eqs.\ (\ref{fiber_mode}) and (\ref{coeff_av}).

\section{Dressed Fiber Mode via EIT}

	In the previous section we have shown that the evanescent-like part of the fiber mode can be coupled with the medium outside the fiber for suitable sets of field, fiber and medium parameters. And thus if the medium outside the thin fiber is active, the fiber mode will closely follow the nature of the active medium. In this section, we demonstrate how a fiber mode can be tailored using an EIT medium surrounding the thin fiber.

	Let us consider that the active medium consists of low density homogeneously broadened atoms / molecules having three-level $\Lambda$ schemes for their energy levels (see the inset of Fig.\ \ref{Fig1}). We assume the atoms / molecules do not move and hence we ignore the transit time broadening. The $|1\rangle \leftrightarrow |2\rangle$ and $|1\rangle \leftrightarrow |3\rangle$ are coupled by a strong control field $\vec{E}_c (r)$ and a weak probe field $\vec{E}_p (r)$ respectively. The control field fiber mode (CFM) and probe field fiber mode (PFM) are given by Eq.\ (\ref{pulse}) with $\alpha \rightarrow c$ for the control field and $\alpha \rightarrow p$ for the probe field.  The $|2\rangle \leftrightarrow |3\rangle$ is a dipole forbidden transition. We assume that both PFM and CFM satisfy the single-mode condition (\ref{single_mode}). The refractive index of a generalized EIT medium, corresponding to a weak probe field is well known \cite{EIT}. Therefore in the following, we only present the formula for refractive index in our configuration, where the constant control field Rabi frequency in \cite{EIT} is replaced by the transversely distributed function $2G(r) = 2d{\cal E}_c(r)/\hbar$; where $d$ is the induced dipole moment corresponding to $|1\rangle \leftrightarrow |2\rangle$ transition, $\hbar$ is Planck constant. Refractive index of such a medium experienced by the weak PFM ${\cal E}_p$ in presence of the strong CFM ${\cal E}_c(r)$ can be obtained as
\begin{equation}
n_m (r) \simeq 1 + \frac{\xi}{2} {\rm Re} 
\left[ 
\frac{i\gamma_1 [\Gamma - i(\Delta -\delta)]}{  
(\gamma_1+\gamma_2+i\delta )(\Gamma -i(\Delta -\delta ))+|G (r)|^2} 
\right],
\label{RI_3lev}
\end{equation}
for small absorption. 
In deriving Eq.\ (\ref{RI_3lev}) we have used following assumptions: (a) the two field modes are uncoupled and do not interact with each other during the propagation inside the fiber, (b)
the CFM ${\cal E}_c$ propagates without loss inside the fiber. We will present a more detailed derivation of the refractive index, for our configuration, using a specific example in Sec.\ V. It may be noted that the medium contribution to the refractive index for the control field is negligibly small as the control field acts on $|1\rangle \leftrightarrow |2\rangle$ transition where very small population or no population is available to make any significant contribution to the susceptibility, and hence the refractive index. Thus for CFM propagation $\bar{n}_m  = n_m= 1$, using vacuum as the medium outside the thin fiber. This $\bar{n}_m$ is used to calculate the fiber mode for the control field ${\cal E}_c$. Here the detuning of the central frequency of PFM (CFM) from the transition frequency $\omega_0$ of $|1\rangle \leftrightarrow |2\rangle$ ($\omega_0^\prime$ of $|1\rangle \leftrightarrow |3\rangle$) is given by $\delta = \omega_{0} - \omega_p$ ($\Delta = \omega_{0}^\prime - \omega_c$); $2\gamma_i$ represents the spontaneous decay coefficient corresponding to $|1\rangle \rightarrow |i\rangle$ transition and $2\Gamma$ represents the dephasing rate of the ground state. It may be noted that for our configuration in Fig.\ \ref{Fig1}, the control field Rabi frequency $2G \equiv 2G(r)$ and hence the refractive index $n_m \equiv n_m(r)$ are functions of $r$ which is different from standard EIT configurations \cite{EIT}. In Eq.\ (\ref{RI_3lev}) we have used a characteristic (dimensionless) parameter $\xi$ for the EIT medium that is defined as
\begin{equation}
\xi = \frac{N|d|^2}{\hbar\epsilon_0 \gamma_1}.
\end{equation}
Here $N$ is the density of atoms / molecules. For typical EIT experiments with alkali atomic gases (see references in \cite{EIT}) $\xi = 0.01 \sim 0.1$. More detailed calculation on a specific realistic system is presented in Sec. V. 

	Assuming the susceptibility of the medium $\chi_m$ to be small and denoting $\bar n$ as the average refractive index over the whole transverse cross section of the fiber,  the propagation constant $\beta_p$ for the fiber mode of the probe field can be approximated as
\begin{eqnarray}
\beta_p &\simeq& \bar{n}k_p 
= \left\{ \frac{\int_{0}^R {\cal E}_p^* n {\cal E}_p r dr}
{\int_{0}^R |{\cal E}_p|^2 r dr} \right\} k_p
\nonumber \\
&=& \left\{b \bar{n}_m + (1-b) n_f \right\} k_p.
\label{beta_approx} 
\end{eqnarray}
Here $\bar{n}_m$, defined in Eq.\ (\ref{n_av}), can be approximately written as
\begin{equation}
\bar{n}_m = \frac{\int_{a}^R {\cal E}_p^* n_m {\cal E}_p rdr}
{\int_{a}^R |{\cal E}_p|^2 rdr},
\label{n_m_bar}
\end{equation}
and the coefficient $b$ denotes the fraction of total energy of the fiber mode outside the fiber and is given as
\begin{equation}
b = \frac{\int_a^R |{\cal E}_p|^2 rdr}{\int_0^R |{\cal E}_p|^2 rdr}.
\label{def_b}
\end{equation}
Clearly from Eq.\ (\ref{beta_approx}), $\beta_p$ of the PFM will be modified by the medium for any non-zero value of $b$. Next we proceed to calculate $b$. 

\begin{figure}
\includegraphics{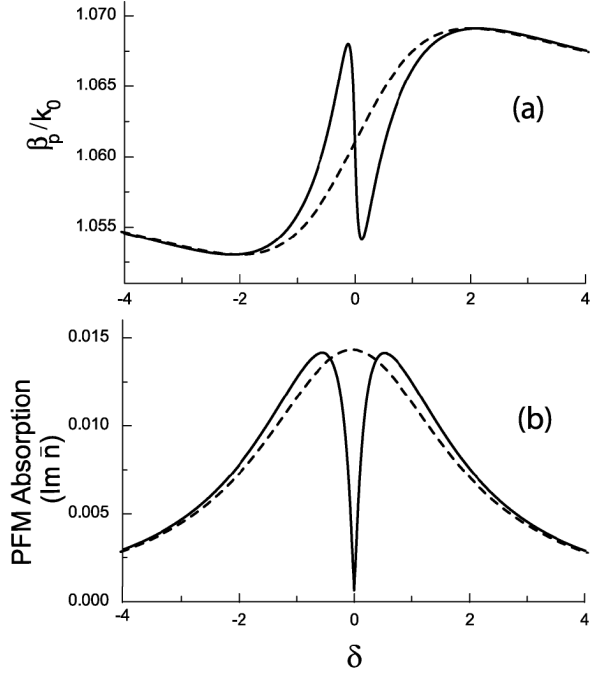}
\caption{\label{Fig2}
(a) The dispersion and (b) the absorption characteristics of the PFM. The dashed lines correspond to the behavior of the PFM alone, and the solid lines depict the modification occured in presence of the CFM (i.e. due to EIT). 
Clearly the behavior of the PFM is governed by the properties of the EIT medium. 
The parameters used here are: $\xi =0.107$, $\gamma_1=\gamma_2=\gamma=1MHz$, $\Gamma=0$, $n_f = 1.43$, $2a = 0.3\mu m$, $\lambda = 780nm$, ${\cal E}_c (r =0) =\hbar\gamma/d$ and $\Delta = 0$. All frequencies are scaled with $\gamma$.
}
\end{figure}

	The distribution for  ${\cal E}_p (r)$, for $r \le a$, is governed by Bessel function $J_0 (\kappa_{f_p}r)$ as in Eq.\ (\ref{fib_modeA}). We expand $J_0(\kappa_{f_p}r) = 1 - \frac{\kappa_{f_p}^2r^2/4}{(1! )^2} + \frac{(\kappa_{f_p}^2r^2/4)^2}{(2! )^2} - ...$ and consider only the first two terms in the expansion for the thin fiber -- where $\kappa_{f_p}r$ is small. Thus substituting ${\cal E}_p$ from Eq.\ (\ref{fiber_mode}) in (\ref{def_b}) and carrying out the integration we obtain
\begin{equation}
b = \left[ 1 + \frac{8\phi_p^2}{3\kappa_{f_p}^2(1 + 2\phi_p a )}
\left( \frac{1}{(1 - \kappa_{f_p}^2 a^2/4)^2} +  \frac{\kappa_{f_p}^2 a^2}{4} -1 \right)
\right]^{-1},
\end{equation}
where $\phi_p$ and $\kappa_{f_p}$ are the coefficients defined in Eq.\ (\ref{coeff_av}). Here we consider the extent of the medium in the radial direction is very large, i.e. $R \rightarrow \infty$. It can be easily shown that $b = 1$ for $a \rightarrow 0$ (and hence $\phi_p a,~\kappa_{f_p} a \rightarrow 0$). Therefore the smaller is the radius $a$, larger is the effect of the surrounding medium on $\beta_p$. For the parameters used in Fig.\ \ref{Fig2}, numerical value of $b \sim 0.57$.

	In Fig.\ \ref{Fig2} we present the dispersion and absorption characteristics of the new modified PFM. In Fig.\ \ref{Fig2}(a), we plot the solution $\beta_p$ (scaled with $k_0$) of the characteristic equation Eq.\ (\ref{beta_eq}). The numerical method for the plot is described in Sec.\ VI. Clearly the PFM assumes the same structure of dispersion and absorption as that of the EIT medium and $\beta_p \equiv \beta_p (G,\Delta ,\delta )$. The dashed lines in Fig.\ \ref{Fig2} represent dispersion and absorption characteristic of the weak PFM when CFM is absent. In presence of the CFM, the absorption of PFM becomes very small at $\delta = 0$. We name this EIT modified PFM with propagation constant $\beta_p$ as {\em dressed} fiber mode. The origin of this new behavior of the fiber mode can be understood as {\em due to the strong coupling of the evanescent-like part of the fiber modes with the surrounding EIT medium}. This is the key for the next sections and for this paper which helps us realize slow light in the thin optical fiber.

\section{Slow Group Velocity of The Fiber Mode}

	In this section we examine the group velocity of the propagating new dressed fiber mode described in the previous section. For PFM with central frequency $\omega_p = \omega_0$ which is on resonance with the $|1\rangle \leftrightarrow |3\rangle$ transition and propagating in the single-mode optical fiber, the group velocity in Eq.\ (\ref{vg_def}) can be rewritten as
\begin{equation}
v_g = \left[ \frac{d\beta_p}{d\omega_p}\right]^{-1}\Bigg|_{\omega_p \approx \omega_0}.
\label{v_g_def}
\end{equation}
It is now clear from Fig.\ \ref{Fig2}(a) that the steep dispersion of $\beta_p$ around $\omega_0$ can cause a large reduction in group velocity.

	To obtain a clear physical picture of the contributions from various dispersions to the group velocity of the fiber mode, we use the approximate relation (\ref{beta_approx}) and substituting into Eq.\ (\ref{v_g_def}) we get 
\begin{equation}
\frac{1}{v_g} =
\frac{\omega_0}{c} \frac{\partial}{\partial \omega_p}\left.\left(
b \bar{n}_m + (1-b) n_f
\right)\right|_{\omega_p \approx \omega_0}
\end{equation}
Dropping the dispersion associated with the fiber-clad [$\partial n_f/\partial \omega_p$] that is negligible compared to that of the resonant medium [$\partial n_m/ \partial \omega_p$] and substituting $\bar{n}_m$ from Eq.\ (\ref{n_m_bar}) we get
\begin{widetext}
\begin{eqnarray}
\frac{1}{v_g} = 
\frac{\omega_0}{c} \left[ (\bar{n}_m -n_f)\frac{\partial b}{\partial \omega_p}
+ \frac{b}{\int_{a}^R {\cal E}_p^2 r dr}
\left(
\int_{a}^R \frac{\partial n_m}{\partial \omega_p} {\cal E}_p^2 r dr
+ 2 \int_{a}^R (n_m - \bar{n}_m) {\cal E}_p\frac{\partial {\cal E}_p}{\partial\omega_p} r dr 
\right)\right]\Bigg|_{\omega_p \approx \omega_0}.
\label{tmp_vg}
\end{eqnarray}
\end{widetext}
In deriving the above equation ${\cal E}_p$ is taken to be real. We have also dropped the terms associated with $\partial k_p/\partial \omega_p$ that are negligibly small ($\sim 1/c$). From Eq.\ (\ref{RI_3lev}) we get 
\begin{equation}
\left. \frac{\partial n_m}{\partial \omega_p}\right|_{\omega_p \approx \omega_0} = 
\frac{\gamma_1\xi}{2}
\frac{|G(r)|^2 - \Gamma^2}{\left[ |G(r)|^2 + (\gamma_1 +\gamma_2)\Gamma\right]^2}
\label{RI_derivative}
\end{equation}
at the EIT condition and when the control field CFM is on resonance with $|1\rangle \leftrightarrow |2\rangle$ transition.
Further using (\ref{fib_modeB}) for ${\cal E}_p(r)$ and ${\cal E}_c (r)$ and using (\ref{RI_derivative}), we obtain an expression for the group velocity of the PFM in the presence of the CFM
\begin{eqnarray}
\frac{1}{v_g} &\simeq&  
\frac{\omega_0\gamma_1\xi}{2c G_0^2} 
\frac{b\phi_p^2\{ 1+2(\phi_p - \phi_c )a\}}{(\phi_p -\phi_c)^2(1+2\phi_p a)}
\nonumber \\
&&- \frac{\omega_0}{c}(n_f -\bar{n}_m)\frac{\partial b}{\partial \omega_p}\bigg|_{\omega_p \approx \omega_0},
\label{vg_final}
\end{eqnarray}
where $G_0 = G (r = a)$. We have considered that the medium is present over large radial distances ($R \rightarrow \infty$). The parameters $\phi_p$ and $\phi_c$ are the PFM and CFM parameters defined in Eq.\ (\ref{coeff_av}) with $\alpha \rightarrow p$ and $\alpha \rightarrow c$ respectively.
We note that in deriving (\ref{vg_final}) we have neglected the third term in (\ref{tmp_vg}) compared to the other two terms, as it contains the small multiplicative term $(\bar{n}_m - n_m)$. Numerically we observed that the third term  is at least 4 to 5 orders of magnitude smaller compared to the other two in (\ref{tmp_vg}), for a thin fiber surrounded by a typical EIT medium. We have derived the above expression for group velocity at the ideal EIT condition, i.e. for $\Gamma = 0$. It is interesting to note from Eq.\ (\ref{vg_final}) that, assuming the second term is small compared to the first, the fiber mode can be brought to a complete halt ($v_g = 0$) by turning off the CFM ($G_0 = 0$), which could be useful for storage \cite{store_QL} of light information in the PFM. Such dynamical aspects of the group velocity of the fiber mode will be discussed elsewhere. 

	Further we examine the limiting case of vanishing fiber radius ($a \rightarrow 0$), where $\kappa_{m_c} \rightarrow 0$ (as $\beta_c \approx k_c$, $\bar{n}_m = n_m = 1$) and hence $\phi_c \rightarrow 0$. Also $b \rightarrow 1$ as discussed in the previous section and hence $\partial b/\partial \omega_p \approx 0$. Substituting these in Eq.\ (\ref{vg_final}) we get
\begin{equation}
\frac{1}{v_g} \simeq \frac{\omega_0\gamma_1\xi}{2c}\frac{1}{G_0^2}.
\label{vg_f1}
\end{equation}
Thus we recover the expression for group velocity of a pulse propagating inside a bulk sample of the EIT medium with the zero ground state dephasing, in the case of vanishing fiber radius.

\section {Ortho-$H_2$ Doped in Solid Para-$H_2$ Crystal As An EIT medium}

\begin{figure}
\includegraphics{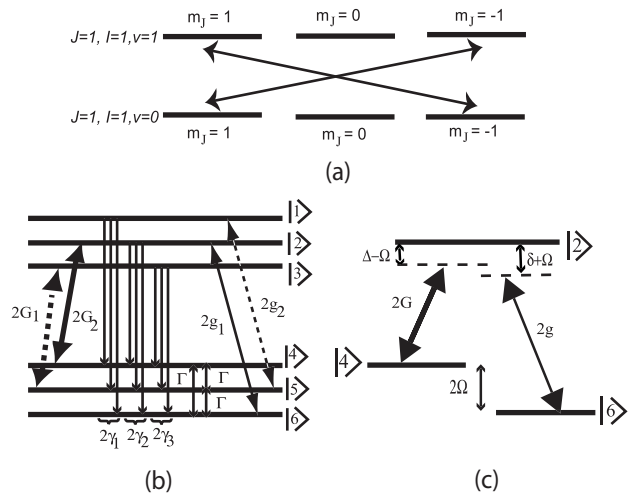}
\caption{
\label{Fig3}
(a) The crystal field induced $v = 0\leftrightarrow v=1$ transition. The selection rule for this transition is $\Delta m_J = \pm2$. (b) The six level scheme that can be obtained from the hyperfine splitting of $I=1$ degenerate levels applying an axial magnetic field. Here $2\gamma_i$ are the non-radiative decay rates from $v=1 \rightarrow v=0$ states, $\Gamma$ is the crystal field induced mixing rate of the ground states, $2G_i$ and $2g_i$ are CFM and PFM Rabi frequencies defined in Eqs.\ (\ref{Rabi_c}) and (\ref{Rabi_p}), respectively. (c) The effective three-level $\Lambda$ scheme reduced from the 6-level scheme which describes the EIT in the ortho-H$_2$ molecule.
}
\end{figure}

We propose the orthohydrogen molecules (with nuclear spin $I =1$) doped in the solid parahydrogen ($I=0$) molecular crystal as an EIT medium. The transition under consideration is a crystal field induced vibrational-rotational transition \textit{Q}$_1$(1) ($v = 1 \leftarrow  0$, $J = 1 \leftarrow  1$, $M_I$ = $-1 \leftarrow  0$, $0 \leftarrow  1$). The degenerate sublevels of rotational angular momentum $J = 1$ states with $m_J = 0, \pm 1$ split in the anisotropic crystal field. Further the quadrapole moment of the orthohydrogen ($J=1$) molecule induce dipole moments on the surrounding parahydrogen ($J=0$) molecules. And these many body dipole moments interact with a radiation, having its electric field perpendicular to $c$-axis of the crystal, to cause the transition $v=0 \leftrightarrow v=1$ \cite{abspn_coeff,kranendonk,oka}. This transition occurs at infrared $\lambda_0 = 2.4 \mu$m and the selection rule for this transition is $\Delta m_J = \pm 2$ [See Fig.\ \ref{Fig3}(a)]. Further applying an axial magnetic field, the degeneracy in the hyperfine levels $I=1$ can be removed to obtain a six level scheme as depicted in the Fig.\ \ref{Fig3}(b). 

	Before we proceed, we note that the selection of this medium as the EIT medium is of particular importance to the present problem, due to the following: (a) As discussed earlier we need the field to satisfy single mode condition (\ref{single_mode}) and also a large evanescent-like part of PFM should be available for interaction of the fiber mode with the medium, for which $2a < \lambda_0$. Thus for ortho-H$_2$ medium ($\lambda_0 = 2.4\mu$m) one can work with $2a \sim 1-2\mu$m - whereas for a standard EIT medium having optical transitions, the thickness of the fiber has to be as small as $2a \ll 1\mu$m. (b) This solid has very narrow widths \cite{oka} compared to the other solid EIT medium \cite{slow_solid}.
(c) Further we note that since we are working here with the solid medium, the assumption of neglecting the transit time broadening in Sec.\ III is easily met. 

	The magnetic degenerate sub-levels in Fig.\ \ref{Fig3}(a) can be coupled selectively by choosing suitable polarizations of the laser fields \cite{mag_degen}. We use a strong $\sigma_+$ polarized laser field that couples $|2\rangle \leftrightarrow |4\rangle$ and $|3\rangle \leftrightarrow |5\rangle$ transitions, and a $\sigma_-$ polarized weak probe field that couples the $|1\rangle \leftrightarrow |5\rangle$ and $|2\rangle \leftrightarrow |6\rangle$ transitions. Thus the interaction Hamiltonian in Rotating Wave Approximation (RWA) can be written as
\begin{eqnarray}
{\cal H}_I &=& -\hbar \big[ (g_1 |2\rangle\langle 6| +g_2|1\rangle\langle 5|)e^{-i\omega_p t +i\beta_p z}
\nonumber \\ 
&& + (G_1 |3\rangle\langle 5| +G_2|2\rangle\langle 4|)e^{-i\omega_c t +i\beta_c z}
+ {\rm H.c.}\big].
\label{H_I}
\end{eqnarray}
Here $\omega_p ~(\omega_c)$ represents the central frequency of the PFM (CFM), and $\beta_p~(\beta_c)$ represents the propagation constant of the corresponding propagating mode. The Rabi frequencies of the CFM are,
\begin{equation}
G_1 = \frac{|\vec{d}_{35}|{\cal E}_c}{\hbar},~
G_2 = -\frac{|\vec{d}_{24}|{\cal E}_c}{\hbar},~
\label{Rabi_c}
\end{equation}
and that of the PFM are
\begin{equation}
g_1 = -\frac{|\vec{d}_{26}|{\cal E}_p}{\hbar},~
g_2 = \frac{|\vec{d}_{15}|{\cal E}_p}{\hbar}.
\label{Rabi_p}
\end{equation}
Here $\vec{d}_{ij} = \langle i|\vec{d}|j\rangle$ represents an effective dipole moment corresponding to the crystal field induced $|i\rangle \leftrightarrow |j\rangle$ transition, which is calculated from the experimentally measured value of the integrated absorption coefficient \cite{abspn_coeff}. The unperturbed Hamiltonian in terms of Fig.\ \ref{Fig4}(b) can be written as
\begin{eqnarray}
{\cal H}_0 = \hbar(\omega_v +2\Omega) |1\rangle\langle 1| 
+ \hbar (\omega_v +\Omega) |2\rangle\langle2| 
\nonumber \\
+ \hbar \omega_v |3\rangle\langle 3|
+ 2\hbar \Omega |4\rangle\langle 4|
+ \hbar \Omega |5\rangle\langle 5|.
\end{eqnarray}
Here $\hbar\omega_v$ is the separation between the states $|n\rangle$ and $|n + 3\rangle$ (for $n = 1,2,3$) in Fig.\ \ref{Fig4}(b). The hyperfine splitting of $I=1$ degenerate levels, caused by the axial magnetic field, is represented by $2\Omega$. 

	Thus the equation of motion for the six-level system is given by the density matrix equation
\begin{eqnarray} 
\frac{\partial \rho }{\partial t} = &-&\frac{i}{\hbar } [{\cal H}_0 +{\cal H}_I,\rho ] 
\nonumber \\
&-& \sum_{i = 1}^3 \gamma_i \left\{|i\rangle\langle i|,\rho \right\}_+ 
+ \frac{2}{3}\sum_{i=1}^3\sum_{j=4}^6 \gamma_i \rho_{ii} |j\rangle\langle j|
\nonumber \\
&-& 2\Gamma\Big( \sum_{j=4}^6 \left\{|j\rangle\langle j|,\rho \right\}_+ 
- \sum^6_{i,j=4~(i\ne j)} \rho_{ii} |j\rangle\langle j|\Big). \nonumber \\
\label{master_eq}
\end{eqnarray}
Here $2\gamma_i$s could be the non-radiative decay rates from each of the $v=1 \rightarrow v=0$ states and $2\Gamma$ is the crystal field induced mixing rate of the ground states. The notation $\{~\}_+$ represent the anti-commutators. Thus the equations for the density matrix elements are
\begin{eqnarray}
\dot{\tilde\rho}_{22} &=& -2 \gamma_2 \tilde\rho_{22} + i G_2 \tilde\rho_{42} -i G_2^* \tilde\rho_{24} 
+i g_2 \tilde\rho_{62} -i g_2^* \tilde\rho_{26}
\nonumber \\
\dot{\tilde\rho}_{24}&=& -(\gamma_2+2\Gamma+i(\Delta -\Omega)) \tilde\rho_{24} 
+ i G_2 (\tilde\rho_{44}-\tilde\rho_{22}) + i g_1 \tilde\rho_{64}
\nonumber \\
\dot{\tilde\rho}_{26}&=& -(\gamma_2+2\Gamma+i(\delta +\Omega )) \tilde\rho_{26} 
+ i G_2 \tilde\rho_{46} + i g_1 (\tilde\rho_{66} - \tilde\rho_{22})
\nonumber \\
\dot{\tilde\rho}_{44} &=& -4\Gamma \tilde\rho_{44} + 2 \Gamma( \tilde\rho_{55} +\tilde\rho_{66})
\nonumber \\
&& +\frac{2}{3} \left( \gamma_1 \tilde\rho_{11} + \gamma_2\tilde\rho_{22} +\gamma_3\tilde\rho_{33} \right)
+ i G_2^* \tilde\rho_{24} - i G_2 \tilde\rho_{42}
\nonumber \\
\dot{\tilde\rho}_{46} &=& -(4 \Gamma + i(\delta -\Delta +2\Omega )) \tilde\rho_{46} 
+ i G_2^* \tilde\rho_{26} - i g_1 \tilde\rho_{42}
\nonumber \\
\dot{\tilde\rho}_{66} &=& -4\Gamma \tilde\rho_{66} + 2 \Gamma( \tilde\rho_{44}+\tilde\rho_{55})
\nonumber \\
&&+ \frac{2}{3} \left( \gamma_1 \tilde\rho_{11} + \gamma_2\tilde\rho_{22} +\gamma_3\tilde\rho_{33} \right) 
 + i g_1^* \tilde\rho_{26} - i g_1 \tilde\rho_{62}. \nonumber\\
\label{den_m_eq}
\end{eqnarray}
In writing Eqs.\ (\ref{den_m_eq}) we have eliminated the temporal and spatial rapid oscillations by making the following transformations
\begin{eqnarray}
\rho_{24} &\rightarrow &\tilde{\rho}_{24} \exp[-i\omega_c t+i\beta_c z],~
\nonumber \\
\rho_{26} &\rightarrow& \tilde{\rho}_{26} \exp[-i\omega_p t+i\beta_p z],
\nonumber \\
\rho_{46} &\rightarrow& \tilde{\rho}_{46} \exp[-i(\omega_p -\omega_c) t+i(\beta_p -\beta_c) z],
\nonumber \\
~{\rm and}~
\rho_{ii} &\rightarrow& \tilde{\rho}_{ii}.
\end{eqnarray}
The detunings of central frequencies of CFM and PFM are given by $\Delta = \omega_v - \omega_c$ and $\delta = \omega_v - \omega_p$, respectively. In Eq.\ (\ref{den_m_eq}) we have presented the equations for the relevant density matrix elements only. However in all the calculations presented below we have taken all the six levels into consideration. The conservation law for the population in different energy levels is
\begin{equation}
\sum_{i = 1}^6 \rho_{ii} = 1.
\end{equation}
By the particular choice of the field configuration (see Fig.\ \ref{Fig4}(b)), the strong CFM $G_i(r)$ pumps the population from $|4\rangle$ and $|5\rangle$, and via the non-radiative decays from $|2\rangle$ and $|3\rangle$, a good initial state  $|6\rangle$ can be prepared in the molecules close to the fiber. From (\ref{den_m_eq}) we numerically estimated the steady state value of ground state $\rho_{66}(r) \approx 0.97$ when the CFM Rabi frequency at $r$ is $G {\rm (r)}= \gamma,~\Delta = 0$. Here we have assumed that $\gamma_i=\gamma$ ($i=1,2,3$) and we have used the parameters $2\gamma = 30$ kHz \cite{gamma} and $2\Gamma = 53$ Hz \cite{Gamma_dph}. The Rabi frequencies are assumed as $|G_i| = G$ ($i=1,2$) for simplicity. Thus the configuration in Fig.\ \ref{Fig4}(b) effectively reduces to a three-level $\Lambda$ scheme as shown in Fig.\ \ref{Fig4}(c) in presence of the $\sigma_+$ polarized strong CFM ($G_i \ne 0$), and hence the dynamics is primarily governed by the Eqs. (\ref{den_m_eq}). 

	Assuming that the steady state population of $|6\rangle$ in absence of PFM is equal to unity, we get the solution of the density matrix equation for a weak PFM as
\begin{equation}
\tilde{\rho}_{26}(r) = \frac{i g_1(r) [4\Gamma + i (\delta - \Delta +2\Omega)]}{(\gamma +2\Gamma +i(\delta +\Omega ))(4\Gamma + i(\delta -\Delta +2\Omega)) + |G(r)|^2}.
\label{rho_26}
\end{equation}
The frequency dependent susceptibility of the medium containing orthohydrogen doped in parahydrogen crystal can be written as
\begin{equation}
\chi_m =  \chi_{\rm para} + \xi \sigma_{26},
\label{chi_m_system}
\end{equation}
where $\chi_{\rm para}$ is susceptibility of the background parahydrogen matrix,  $\sigma_{26} \equiv \gamma\tilde{\rho}_{26}/g_1$ and $\xi = Nd_{\rm eff}^2/(\hbar\epsilon_0\gamma)$ (where $d_{\rm eff} = d_{ij}$ effective dipole moment, assuming the effective dipole strengths of all involved transitions to be same). All other parameters are same as defined earlier. Thus the refractive index of this doped medium is
\begin{equation}
n_m (\omega ) = (1+{\rm Re}~\chi_m)^{1/2} = [n^2_{\rm para} + \xi ({\rm Re}~  \sigma_{26} )]^{1/2},
\label{total_RI}
\end{equation}
where $n_{\rm para}$ is the refractive index of the background parahydrogen. 
Therefore the refractive index of the medium averaged over the transverse distribution of the PFM, using the definition in Eq.\ (\ref{n_av}), can be written as
\begin{equation}
\bar{n}_m \equiv \left[ n_{\rm para}^2 + \xi \frac{\int_a^R {\cal E}_p^* (r) [{\rm Re}~\sigma_{26} (r)] {\cal E}_p(r) dr}{\int_a^R |{\cal E}_p(r)|^2 dr}\right]^{1/2}
\label{n_m_ortho}
\end{equation}
However assuming that the susceptibilities $\chi_{\rm ortho}$ and $\chi_{\rm para}$ are small, we can rewrite the above as
\begin{equation}
\bar{n}_m \equiv n_{\rm para} + \frac{\xi}{2}\frac{\int_a^R {\cal E}_p^* (r) [{\rm Re}~\sigma_{26} (r)] {\cal E}_p(r) dr}{\int_a^R |{\cal E}_p(r)|^2 dr}.
\end{equation}
Using this $\bar{n}_m$ and the same procedure as described in Sec. III and IV, we calculate $\beta_p$ and hence the group velocity of the PFM inside the thin fiber 
surrounded by ortho-H$_2$ doped solid para-H$_2$ crystal:
\begin{eqnarray}
\frac{1}{v_g} = 
\frac{\omega_0\gamma\xi}{2c G_0^2} 
\frac{b\phi_p^2\{ 1+2(\phi_p - \phi_c )a\}}{(\phi_p -\phi_c)^2(1+2\phi_p a)}
\nonumber \\
- \frac{\omega_0}{c} (n_f - \bar{n}_m) 
\frac{\partial b}{\partial \omega_p}\bigg|_{\omega_p \approx \omega_0},
\end{eqnarray}
assuming the ground state mixing rate $2\Gamma = 0$. Here the parameters $G_0,~\xi,~b, ~\phi_p,~\phi_c$ are same as defined earlier. For simplicity we have derived the above equation at $\Omega = 0$ and use this condition in all further calculations. This analytical result agrees well in the order of magnitude estimation with the numerical result presented in the next section. 

In the following section we present the numerical results and discussions on various aspects of slow light propagation of dressed mode propagating inside the thin fiber with orthohydrogen doped in parahydrogen crystal as the EIT medium.

\section{Numerical Results and Discussion}

	 The propagation of the probe field inside the thin fiber is numerically carried out using the two-dimensional (2D) {\em beam propagation method} (BPM) \cite{marcuse} that solves the 2D scalar equation
\begin{equation}
\frac{\partial^2 {\cal A}_p}{\partial x^2} 
+ \frac{\partial^2 {\cal A}_p}{\partial z^2} + n^2(x,z)k_p^2{\cal A}_p = 0,
\end{equation}
with $x$ as the only transverse component. Here ${\cal A}_p$ is the amplitude of the probe field; $\vec{E}_p \equiv \hat{{\cal E}} \int_{-\infty}^\infty {\cal A}_p (x;\omega) e^{ik_pz-i\omega t} d\omega +c.c.$ used for initiation of the probe field propagation which eventually evolves to the fiber mode ${\cal E}_p$ shown in Fig.\ 4(a).
The solution to the above equation is obtained by 
discretizing space and assuming the medium to be a combination 
of short sections of homogeneous medium, having a constant refractive
index $\bar{n}$, separated by lenses at the grid points that incorporate the effect of the actual refractive 
index of the medium $n(x,z)$ at the given $(x,z)$. Thus the field propagation
is carried out in two steps: (1) {\em homogeneous step}: The homogeneous part
of the propagation was carried out by Fourier decomposition in terms of 
plane waves; (2) {\em lens step}: In this step a phase shift is introduced to incorporate the effect of $n(x,z)$. 
We have included following features into 2D BPM to describe the dynamics of our system: (a) a complex refractive index $n_m(x)$ is used to account for the absorption in the resonant medium; (b) the averaged refractive index $\bar{n}$ for the homogeneous step is determined adaptively during the propagation of the field using 
\begin{equation}
\bar{n}^2 = \frac{\int_{-R}^R {\cal A}_p^*(x) n^2 (x) {\cal A}_p(x) dx}{\int_{-R}^R |{\cal A}_p (x)|^2 dx};
\label{nbar}
\end{equation}
and (c) the energy conservation $\partial [\int_{-R}^{R}|{\cal A}_p(x)|^2 dx]/\partial z = 0$ is used to compensate the numerical losses in the homogeneous steps.

\begin{figure}
\includegraphics{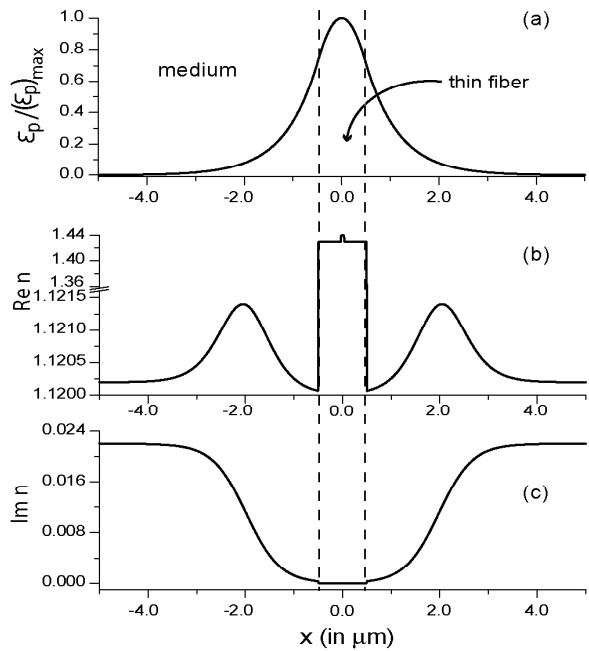}
\caption{
\label{Fig4}
Distribution of the (a) PFM (scaled), (b) the real part of refractive index, and (c) the imaginary part of the refractive index that represent the absorption, along the transverse direction.
}
\end{figure}

\begin{figure}
\includegraphics{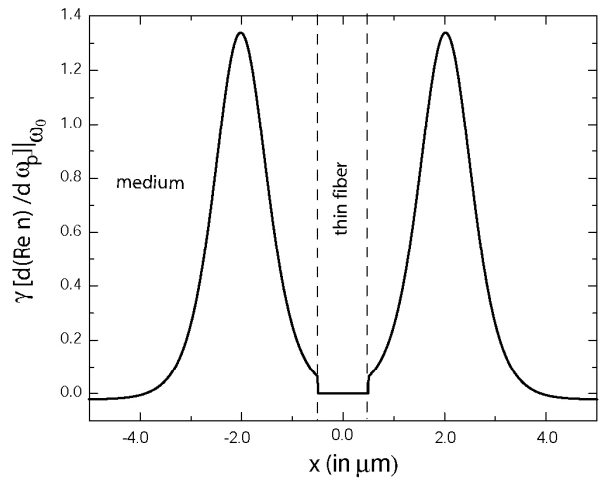}
\caption{
\label{Fig5}
Variation of slope of the real part of the refractive index with respect to the PFM frequency in the transverse direction.
}
\end{figure}

\begin{figure}
\includegraphics{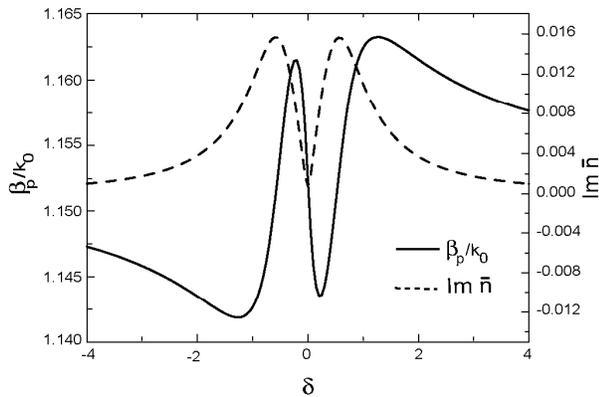}
\caption{
\label{Fig6}
The dispersion (solid line) and the absorption (dotted line) characteristics of the dressed fiber mode when the thin fiber is surrounded by ortho-H$_2$ doped para-H$_2$ crystal as the EIT medium. Here all the parameters are same as in Fig.\ \ref{Fig4}., and the PFM detuning $\delta$ is scaled with $\gamma$.}
\end{figure}

	The numerical evaluation of the dressed fiber mode PFM, shown in Fig.\ \ref{Fig4}(a) involves the following steps. First step is: to determine the transverse distribution of the CFM ${\cal E}_c (x)$ using Eqs.\ (\ref{fiber_mode}) and (\ref{beta_eq}) with $n_m = n_{\rm para} = 1.12$. We use this ${\cal E}_c(x)$ in Eq.\ (\ref{rho_26}) and in all further calculations, and evaluate the response of the medium to the PFM in the presence of the CFM. We initiate the propagation of probe field with a Gaussian field having FWHM $2a$ and numerically propagate using the 2D BPM as described above. In few steps, the Gaussian probe evolves to the PFM ${\cal E}_p (x)$ as shown in Fig.\ \ref{Fig4}(a). The distribution of refractive index and absorption along the transverse direction are shown in Fig.\ \ref{Fig4}(b) and (c) respectively. Clearly the behavior of the PFM is mainly determined by the region close to this fiber (for $|x| < 2~\mu$m), as ${\cal E}(x)$ is very small outside this region. The absorption is large beyond this region as the control field reduces and hence EIT condition does not hold beyond this region. The parameters used here are: $N \approx 1.3\times 10^{27}~m^{-3}$, $d_{\rm eff} \approx 7.3\times 10^{-34}$ Coul.m, $2\gamma = 30$ kHz, $\Gamma = 1.17\times 10^{-3}\gamma$, the resonant $\lambda_0 = 2.4~\mu$m, $\delta = -0.001\gamma$ and $({\cal E}_c)_{max} = \hbar\gamma/d_{\rm eff}$, the corresponding laser intensity 
is $\approx 0.6$ W/cm$^2$. The above value of $N$ corresponds to 5\% doping of ortho-H$_2$ molecule in a matrix of para-H$_2$ crystal. However since the inhomogeneous width in a solid is much larger compared to its natural linewidth, to achieve a good EIT condition, a larger intensity is required to overcome the inhomogeneous width \cite{slow_solid}. Using an effective width of $2\gamma =$ 20.03 MHz -- where the additional $20$ MHz is to account for a possible inhomogeneous broadening ($2\gamma_{\rm inh}$) in the crystal -- the intensity is estimated to be $\approx$279 kW/cm$^2$. This intensity for our configuration corresponds to a laser power of $\approx 19.7$ mW only (assuming the beam diameter of the PFM of the thin fiber to be $\sim$3 $\mu$m). Further some early experimental observations have predicted possibility of a much smaller inhomogeneous width for a hydrogen crystal prepared under very good experimental condition -- that would reduce the power requrement for the control field by square times the the amount by which inhomogeneous width is reduced; e.g. if inhomogeneous width is 20 times smaller, the power required will be reduced by 400 times. Here for the approximate estimation, $\gamma_{\rm inh}$ is assumed to be Lorentzian \cite{inh_broad} and added to the natural linewidth to get the effective $\gamma$. In Fig.\ \ref{Fig5} we show the variation of the slope of the refractive index along the transverse direction. It is clear from Fig. \ref{Fig5} that the slope is positive only in the region close to the fiber, and hence favorable for the slow group velocity. For $|x| >2~\mu$m the slope decreases and becomes negative for $|x| > 4~\mu$m. However, the contribution from the later part is extremely small as the PFM ${\cal E}_p (x)$ has very small value beyond $|x| >2~\mu$m. Further ${\cal E}_p (x)$ is used to determine the complex $\bar{n}_m$ using Eq.\ (\ref{n_m_ortho}) and hence $\beta_p$ could be determined using Eqs.\ (\ref{beta_eq}) and (\ref{coeff_av}). In Fig.\ \ref{Fig6} we plot the scaled $\beta_p$ and Im $\bar{n}_m$ as function of the PFM detuning $\delta$. The corresponding group velocity $v_g$ of the PFM $\approx$44.1 m/sec. The resulting group delay of the PFM is as large as $\approx$1.13 $\mu$sec for a medium length of $50~\mu$m (along the direction of propagation). For completeness we note that the group velocity of a weak probe in free space, passing through the bulk solid sample of orthohydrogen doped in a parahydrogen matrix, is estimated to be $v_g \approx 52.95$ m/sec. We have used a control field strength equal to ${\cal E}_c(r=a)$; all other field and medium parameters are taken to be the same as in the above. It may be noted that the group velocity reduction in a fiber is larger than that in a bulk EIT medium (without fiber), which is due to following reason: group velocity of probe pulse is directly proportional to the control field intensity. In case of a fiber, the control field itensity decreases along the transverse direction causing the tail part of the mode to move extremely slowly, which reduces the group velocity of the fiber mode drastically. However in the case of a bulk sample, the intensity is assumed to be constant along the transverse direction. Therefore with the same peak value for control field, the probe field pulse velocity is slower in the the fiber compared to that in the bulk medium.

	It may further be noted that $\beta_p$ is a constant of propagation only if we assume that the absorption of both pump and probe are negligible. Because a change in pump or probe intensity distribution could change $\bar{n}_m$ and hence $\beta_p$.
Validity of this approximation lies in the fact that we are working
under EIT condition Im $\bar{n}_m \simeq 0$ around $\delta =0$ -- which, however, is good only in the region close to the fiber. Moreover the probe field amplitude drops rapidly outside this region. Hence the probe absorption and dispersion characteristics are primarily determined by the region close to the fiber, where EIT condition is satisfied. 
	
	Apart from the usefulness of our configuration to reduce the group velocity of the fiber mode, it offers following advantages: a fiber mode, of the original (undistorted) fiber with large width at the input -- when propagates through the thin (tapered) region, the width is considerably reduced. Thus the tapered fiber acts like a focusing element and hence a large part the field energy is available in the evanescent-like part of the thin fiber mode -- that could be useful to access weaker atomic/molecular transitions such as the Q$_1 (1)$ transition of ortho-H$_2$. 

\section{Conclusion}
	
	In summary, we have demonstrated that the group velocity of a fiber mode propagating inside a thin single-mode optical fiber can be reduced drastically utilizing the EIT. We have shown that due to the strong interaction of the evanescent-like part of the fiber mode with the EIT medium surrounding the fiber, the fiber mode picks up the dispersion and absorption properties of the medium. We have analyzed in detail how a dressed fiber mode can be prepared and the slow fiber mode propagation can be realized when the fiber is surrounded by an EIT medium, both numerically and with approximate analytical results. We have introduced orthohydrogen doped in parahydrogen crystal as a new solid EIT medium and demonstrated in detail that using this as the medium surrounding the fiber, we can reduce the group velocity of the fiber mode to $\sim$44 m/sec for the chosen set of CFM parameters. From the analytical result we also predict the possibility of stopping the PFM by dynamically switching off the CFM.

\begin{acknowledgements}
We are thankful to Fam Le Kien for many invaluable discussions.
AKP gratefully acknowledges the support from Japanese Society for Promotion of Sciences (JSPS).
\end{acknowledgements}

\end{document}